\documentclass[aps,nofootinbib,superscriptaddress,tightenlines,twocolumn,pra]{revtex4-1}

\usepackage{amsmath,amssymb,amstext,amsthm}
\usepackage[pdftex]{graphicx}
\usepackage{xcolor}
\usepackage[pdftex,letterpaper=true,pagebackref=false]{hyperref}

\hypersetup{
pdftitle={Cavity cooling to the ground state of an ensemble quantum system},
pdfauthor={Christopher J. Wood},
	pdfnewwindow=true, 
	colorlinks=true, 
	linkcolor=blue, 
	citecolor=magenta, 
	urlcolor=blue
}
 
\newcommand{\dket}[1]{\mbox{$\left|\left.#1\right\rangle\!\right\rangle$}}
\newcommand{\dbra}[1]{\mbox{$\left\langle\!\left\langle #1\right.\right|$}}
\newcommand{\dbradket}[2]{\mbox{$\langle\!\langle #1|#2\rangle\!\rangle$}}

\DeclareMathOperator{\Tr}{Tr}

\DeclareMathAlphabet{\bb}{U}{bbold}{m}{n}
\newcommand{\id}{{\bb 1}}
\def\1#1{{\bf #1}}
\def\2#1{{\cal #1}}
\def\3#1{{\sl #1}}
\def\4#1{{\tt #1}}
\def\5#1{{\sf #1}}
\def\6#1{{\mathfrak #1}}
\def\7#1{{\mathbb #1}}

\definecolor{dred}{rgb}{.8,0.2,.2}
\definecolor{ddred}{rgb}{.8,0.5,.5}
\definecolor{dblue}{rgb}{.2,0.2,.8}
 
\begin{document}
\title{Cavity cooling to the ground state of an ensemble quantum system}
\author{Christopher J. Wood}
\email{cjwood@cjwood.com}
\affiliation{Institute for Quantum Computing, University of Waterloo, Waterloo, ON N2L 3G1, Canada}
\affiliation{Department of Physics and Astronomy, University of Waterloo, Waterloo, ON N2L 3G1, Canada}

\author{David G. Cory}
\affiliation{Institute for Quantum Computing, University of Waterloo, Waterloo, ON N2L 3G1, Canada}
\affiliation{Department of Chemistry, University of Waterloo, Waterloo, Ontario N2L 3G1, Canada}
\affiliation{Perimeter Institute for Theoretical Physics, Waterloo, Ontario N2L 2Y5, Canada}
\affiliation{Canadian Institute for Advanced Research, Toronto, Ontario M5G 1Z8, Canada}

\date{March 1, 2016}

\begin{abstract}
We describe a method for initializing an ensemble of qubits in a pure ground state by applying collective cavity cooling techniques in the presence of local dephasing noise on each qubit. To solve the dynamics of the ensemble system we introduce a method for dissipative perturbation theory that applies average Hamiltonian theory in an imaginary-time dissipative interaction frame to find an average effective dissipator for the system dynamics. We use SU(4) algebra generators to analytically solve the first order perturbation for an arbitrary number of qubits in the ensemble. We find that to first order the effective dissipator describes local $T_1$ thermal relaxation to the ground state of each qubit in the ensemble at a rate equal to the collective cavity cooling dissipation rate.
The proposed technique should permit the parallel initialization of high purity states in large ensemble quantum systems based on solid-state spins.
\end{abstract}

\maketitle

\section{Introduction}
A fundamental challenge to implementing quantum information processing on a physical device is the ability to rapidly and repeatable initializing the quantum system in a high purity state. Depending on the physical architecture of a device, quantum state initialization is typically done using methods such as strong projective measurements or filters~\cite{KLM2001}, optical pumping and atomic transitions~\cite{Weber:2010a,Gumann2014}, laser and microwave cooling~\cite{Wineland:1979a,Monroe:1995a,Wallquist:2008a,Valenzuela:2006a}, dynamical nuclear polarization~\cite{AbragamBook,Ramanathan:2008a}, algorithmic cooling~\cite{Baugh2005,Ryan2008}, and dissipative state engineering~\cite{Verstraete:2009,Lin:2013}. As the number of qubits in quantum devices increases initialization methods that can be implemented in parallel across many qubits are necessary to enable scalability. In the case of spin ensembles thermal relaxation processes naturally initialize all spins in parallel and can be used for state preparation, however to achieve high purity state at thermal equilibrium very low temperatures and strong magnetic fields are needed. In the case of solid-state electron spin systems this can be a limiting factor as the thermal relaxation time ($T_1$) becomes very long at low temperatures~\cite{AbragamBook}.

In recent work by the authors and others it was suggested that cavity cooling techniques could enable the collective removal of entropy from an ensemble spin system~\cite{Butler:2011,WoodCC}, and this has recently been experimentally demonstrated~\cite{Bienfait:2015}. Our proposal in \cite{WoodCC} relied on engineering a Tavis-Cummings (TC) exchange interaction between the spin ensemble and the side band of a high-Q resonator that was actively cooled to a low temperature thermal state. This is of particular interest as there have been several recent designs and proposals for the readout and control of electron spin ensembles using high-Q superconducting resonators~\cite{Kubo:2010a,Benningshof:2012a,Malissa2013,Sigillito2014}.
Due to the identical particles in the ensemble the dynamics preserve a global SU(2) symmetry and the state space has a block diagonal structure with each subspace block corresponding to an irreducible representation of SU(2) that behaves as an effective spin-$J$ particle~\cite{Wesenberg2002,Baragiola2010}. The effective interaction experienced by the spin ensemble was dissipative cooling of each subspace to its respective low temperatures state. The largest effective spin subspace has $J=N/2$ and is called the totally symmetric subspace, or Dicke subspace, and contains all states which are permutation invariant across all $N$ spins~\cite{Dicke1954}.
The cavity cooling dissipator derived in \cite{WoodCC} preserves the global SU(2) symmetry, and so does not couple the subspaces. In this case, if cooling from a maximally mixed initial state the initial population in each subspace will be trapped in that subspace's ground state, and the final state will be a mixed state sum of these ground states.
The open problem is how to solve the dynamics of cavity cooling an ensemble quantum system while including an interaction which couples between these subspaces to enable cooling to the true ground state of the ensemble. The challenge is that once such an interaction is included, in principle the full Hilbert space for $N$-spins must be included which has dimension $4^{N}$.

In the present paper we analyze a method for cavity cooling to the ground state of the ensemble by introducing a local dephasing ($T_2$) noise process on each spin in the ensemble. This acts to distinguish each spin and thus breaks the collective symmetry of the ensemble. In practice dephasing is a phenomenological description of noise in solid-state spin systems can be caused by inhomogeneous broadening, or spurious coupling to neighbouring spins or a local spin bath~\cite{AbragamBook}.
The effects of local dissipation on collective dynamics has been considered previously, where it was shown to rapidly decohere coherent states in the Dicke subspace~\cite{Baragiola2010}. Decoherence of the Dicke subspace has also been studied for an inhomogeneouly broadened ensemble of qubits coupled to a cavity~\cite{Kurucz:2011}. Unlike these previous studies we are able to use local dissipation as a resource for dissipative quantum state engineering to the ground state of the ensemble.

Similar theoretical results to have been presented for cooling an ensemble of nuclear spins by coupling to the motion of a nanoscale mechanical resonator~\cite{Butler:2011}. It was shown that the addition of a chemical shift to each spin was in principle sufficient to break the symmetry and achieve exponential relaxation to the ground state and this was numerically demonstrated for five spins. A similar result was found numerically in \cite{Woggon:2005} where the relaxation rates of the Dicke and non-Dicke subspaces were simulated for 10 inhomogeneously broadened qubits. 
Our work differs from the approaches in \cite{Woggon:2005,Butler:2011} by including the symmetry breaking mechanism as a dissipative term leading to a Lindblad master equation that we can solve perturbatively to derive an analytic expression for the cooling dynamics in the regime where the first order perturbation term dominates the dynamics.
In order to solve the master equation we develop a perturbation theory technique for dissipative evolution. This involves applying the Magnus expansion~\cite{Magnus1954,Blanes2009}, or Average Hamiltonian Theory~\cite{Haeberlen1968,Haeberlen1976,Ernst1987}, in an imaginary-time dissipative interaction frame to the superoperators describing evolution. There is a long history of applying average Hamiltonian theory to superoperators, called Average Liouvillian Theory~\cite{Levitt1992,Ghose1999,Ghose2000}, and also the related cumulant expansion approach for stochastic noise processes~\cite{Kubo1962,Cappellaro2006}. In both of these cases the relevant interaction frame is defined by a Hamiltonian, which typically corresponds to a sequence of control pulses, and the average affect of a dissipative term in this frame is assessed. 
Our approach extends these formalisms by providing a procedure to apply these techniques in a purely dissipative (non-periodic) interaction frame. The utility of this method is that in a system with multiple decoherence mechanisms one can find the average effective dissipation of one mechanism in the presence of another. As we demonstrate, this may then be used for finding the equilibrium state of the system for dissipative state engineering applications.

To analytically solve the lowest order term in the Magnus expansion we use recently introduced techniques for describing the local superoperators on spin ensembles using SU(4) algebra generators~\cite{Hartmann2012,Minghui2013}. In this representation the dynamics preserve a global SU(4) symmetry and any thermal state of the Hamiltonian with SU(2) symmetry will be entirely contained in the totally symmetric subspace of SU(4). 
By considering the explicit matrix representation of the $N$ qubit totally symmetric subspace of SU(4), this approach allows us to numerically simulate the reduced dynamics of the spin ensemble for up to $N=100$ spins on a desktop computer. 

\section{Cavity cooling with local dephasing.}
Consider an ensemble of $N$ identical spin 1/2 particles interacting with a single mode cavity. Let $a, a^\dagger$ be the cavity lowering and raising operators respectively, and 
\begin{equation}
J_\alpha = \sum_{j=1}^N \frac{\sigma_\alpha^{(j)}}{2}, \quad J_\pm = \sum_{j=1}^N \sigma_\pm^{(j)},
\end{equation}
be the collective spin operators for the spin ensemble, where $\sigma_\alpha^{(j)}, \alpha=x,y,z$ is the Pauli matrix for the $j^{th}$ spin-$\frac12$ system. 
If the spins are on-resonance with the cavity, or are driven to be on-resonance with a side-band of the cavity as was considered in \cite{WoodCC}, then the spin-cavity interaction is well described by the TC Hamiltonian 
\begin{equation}
H_{\text{TC}} = g(J_+ a + J_- a^\dagger), 
\end{equation}
which is the $N$ spin generalization of the Jaynes-Cummings interaction~\cite{Tavis1968,Tavis1969}. The cavity will experience photon-loss at a rate inversely proportional to $Q$, which may be described by the Lindblad dissipator 
\begin{equation}
\2D_c = \kappa(1+\overline{n})D[a]+ \kappa\overline{n}D[a^\dagger],
\end{equation}
where 
\begin{equation}
D[a]\rho = a\rho a^\dagger-\frac12\{a^\dagger a,\rho\}.
\end{equation}
The effect of $\2D_c$ is to reset the cavity to a thermal state $\rho_{\text{eq}}$ satisfying $\overline{n} = \Tr[a^\dagger a\,\rho_{\text{eq}}]$, at a rate $\kappa$. 
In this case the evolution of the joint spin-cavity system is described by the Lindblad master equation
\begin{equation}
\frac{d}{dt}\rho(t) = -i [H_{\text{TC}},\rho(t)] + \2D_c\, \rho(t).
\label{eq:tc-me}
\end{equation}
It was shown in \cite{WoodCC} that by adiabatically eliminating the cavity in the Markovian regime ($\kappa \gg g\sqrt{N}$), this interaction leads to an effective spin-dissipation of the form 
\begin{equation}
\frac{d}{dt}\rho_s(t) = \2D_{cc}\rho_s(t),
\end{equation}
where $\rho_s$ is the density matrix of the spin ensemble alone after adiabatic elimination of the cavity, and
\begin{equation}
\2D_{cc} = \Gamma\, (1+\overline{n})D[J_-] + \Gamma\, \overline{n} D[J_+],
\label{eq:cc-diss}
\end{equation}
is the cavity cooling dissipator with rate $\Gamma = 4g^2/\kappa$. 
For $\overline{n}=0$, the dissipator $\2D_{cc}$ cools each of SU(2) irrep subspaces of the collective spin ensemble to their respective ground states. However, since $\2D_{cc}$ preserves the SU(2) symmetry it does not couple the subspaces. Thus if the system is initially in a highly thermal state, cavity cooling dissipator alone is not sufficient to cool to the ground state of the system. 

To enable cooling to the ground state we need to introduce an interaction that breaks the SU(2) symmetry of the ensemble, yet does not inhibit the cavity cooling dissipator. This may be achieved by a local dephasing term which acts identically on each spin in the ensemble. Dephasing, or $T_2$ dissipation, of a single spin-half system causes the off-diagonal density matrix elements of the spin to decay exponentially to zero. It is generated by the Lindblad dissipator
\begin{equation}
D[\sigma_z/2]\rho \equiv \frac14(\sigma_z\rho \sigma_z^\dagger - \rho).
\end{equation}
The dissipator for identical local dephasing dissipators on each spin in the ensemble is then given by 
\begin{equation}
\2D_{T_2} = \sum_{j=1}^N \gamma D[\sigma_z^{(j)}/2]. 
\end{equation}
This process results in a mixing of states across different spin-$J$ subspaces that have the same $J_z$ value. Since cavity cooling drives each subspace to the lowest $J_z$ value state, a $T_2$ process will leak population trapped in the ground state of a spin-$J$ subspace into the $k^{\text{th}}$ excited state of a spin-$(J+k)$ subspace, which will then be cooled to that subspaces ground state. In the ideal case this eventually leads to the ground state of the Dicke subspace. To show this we must solving the the dynamics of the spin master equation
\begin{equation}
\frac{d}{dt}\rho_s(t) = (\2D_{cc} + \2D_{T_2})\rho_s(t).
\label{eq:cct2}
\end{equation}

\subsection{Perturbative Solution}
We now introduce a new dissipative perturbation theory approach to solve Eq.~\eqref{eq:cct2}. To begin we move into a dissipative interaction frame defined by the dephasing dissipator $\2D_{T_2}$. In this interaction frame we have
\begin{equation}
\frac{d}{dt}\widetilde{\rho}_s(t) = \widetilde{\2D}_{cc}(t)\,\widetilde{\rho_s}(t),
\end{equation}
where 
\begin{align}
\widetilde{\rho}_s(t) =& e^{t \2D_{T_2}}\rho_s(t),	\\
\widetilde{\2D}_{cc}(t) =&  e^{t \2D_{T_2}}\2D_{cc}e^{-t \2D_{T_2}}.
\end{align}
Since this is a dissipative interaction frame, the time dependent terms in $\widetilde{\rho}_s(t), \widetilde{\2D}_{cc}(t)$ will be of the form $e^{\pm \omega t}$ for real parameters $\omega$. Hence the terms $e^{+\omega t}$ diverge as $t$ increases.  By transforming into imaginary time ($t\mapsto i\tau$) we may convert this dissipative interaction frame into a periodic one leading to the periodic differential equation
\begin{equation}
\frac{d}{d\tau}Q(\tau) = i \,G(\tau)\,Q(\tau),
\label{eq:wick-me}
\end{equation}
where 
\begin{equation}
Q(\tau) \equiv \widetilde{\rho}(i\tau),\quad G(\tau) \equiv \widetilde{\2D}_{cc}(i\tau).
\end{equation}
The explicit time dependence of the operator $G(\tau)$ is given by
\begin{align}
G(\tau) &= \Gamma(1+\overline{n})G_-(\tau) + \Gamma \overline{n}G_+(\tau) \\
G_\pm(\tau) &= D[J_\pm] 
	+ (e^{\pm i \gamma\tau}-1)A_\pm 	+ (e^{\mp i \gamma\tau}-1)B_\pm
\end{align}
where $\overline{n}$ is the average thermal photon number of the cavity, and $A_\pm, B_\pm$ may be expressed in terms of SU(4) generators (See Appendix~\ref{app:su4}). Hence $G(\tau)$ is periodic with period $T=2\pi/\gamma$, where $\gamma$ is the single spin dephasing rate.
Eq.~\eqref{eq:wick-me} has the formal solution 
\begin{equation}
Q(\tau) = \2T \exp\left(i \int_0^\tau ds G(s)\right)Q(0),
\label{eq:wick-sol}
\end{equation}
where $\2T$ is the time-ordering operator. If the dephasing rate is strong, then we can consider a perturbative expansion of Eq.~\eqref{eq:wick-sol} in terms of the Magnus expansion~\cite{Magnus1954} to find an average time-independent dissipator over the period $T$: 
\begin{equation}
 \2T \exp\left(i \int_0^T ds G(s)\right) = 
	\exp\left(i T \sum_{k=1}^\infty \overline{D}_k\right).
\end{equation}
Since the average dissipator is time-independent, we may transform back into real time to obtain an average description of the system dynamics  over the period $T$, which may then be used to compute the stroboscopic evolution over integer multiples of the period $T$:
\begin{equation}
\widetilde{\rho}_s(nT) = \exp\left({n T \sum_{k=1}^\infty \overline{D}_k}\right)\widetilde{\rho}_s(0).
\end{equation}
If the dephasing rate is greater than the collective cavity cooling rate we may make a secular approximation and only consider the lowest order term in the average dissipator Magnus expansion: 
\begin{equation}
\overline{D}_1 = T^{-1}\int_0^T ds G(s).
\end{equation}
The first order expansion of Eq.~\eqref{eq:cct2} is then given by 
\begin{equation}
\rho_s(t) = e^{t\overline{\2D}_1}\widetilde{\rho}_s(t),
\end{equation}
where
\begin{align}
\overline{\2D}_1 &= 
	\Gamma(1+\overline{n})\, \overline{G}_-
	+ \Gamma\overline{n}\, \overline{G}_+,
	\label{eq:1st-diss}	\\
\overline{G}_\pm &= D[J_\pm] -A_\pm-B_\pm.
\end{align}

For evolution under the first order average dissipator $\overline{D}_1$ in Eq.~\eqref{eq:1st-diss} the expectation value of the ensemble magnetization is given by (See Appendix~\ref{app:localt1})
\begin{equation}
\langle J_z(t)\rangle
	=  e^{-t/ T_1}\langle J_z(0)\rangle
	-\frac{N\left(1-e^{-t/T_1}\right)}{2+4\overline{n}}.
	\label{eq:jz-exp}
\end{equation}
where the relaxation time constant is given by 
\begin{equation}
T_1 = \frac{1}{\Gamma(1+2\overline{n})}.
\end{equation}
This is an exponential relaxation process to an equilibrium state with magnetization
\begin{equation}
\langle J_z\rangle_{eq} = -\frac{N}{2+4\overline{n}}.
\end{equation}
Thus in the ideal cooling limit ($\overline{n}=0$) this is a $T_1$ process to the ground state of the spin ensemble with magnetization $\langle J_z\rangle_{eq} = -N/2$, and $T_1=\Gamma^{-1}$.

\subsection{Simulation}\label{sec:sim}
\begin{figure*}[htbp]
\begin{center}
\includegraphics[width=0.9\textwidth]{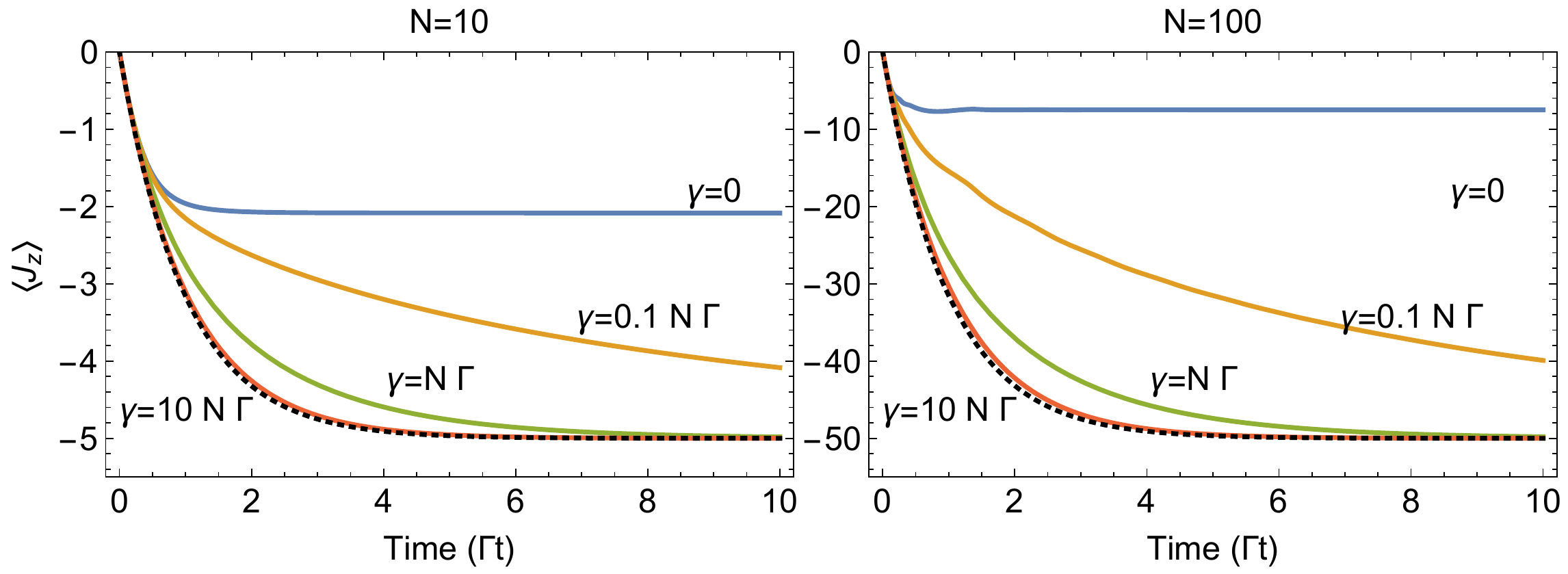}
\caption{Simulations of the magnetization expectation value $\langle J_z(t)\rangle$ for a maximally mixed initial state of an ensemble of $N=10$ spins (left), and $N=100$ spins (right) for master equation described by first order average dissipator in Eq.~\eqref{eq:1st-diss} (dotted black line), and of the cavity cooling with local dephasing master equation in Eq.~\eqref{eq:cct2} for $\gamma = \lambda N \Gamma$, and $\lambda = 0, 0.1, 1, 10$. 
}
\label{fig1}
\end{center}
\end{figure*}

\begin{figure*}[htbp]
\begin{center}
\vspace{-1em}
\includegraphics[width=0.9\textwidth]{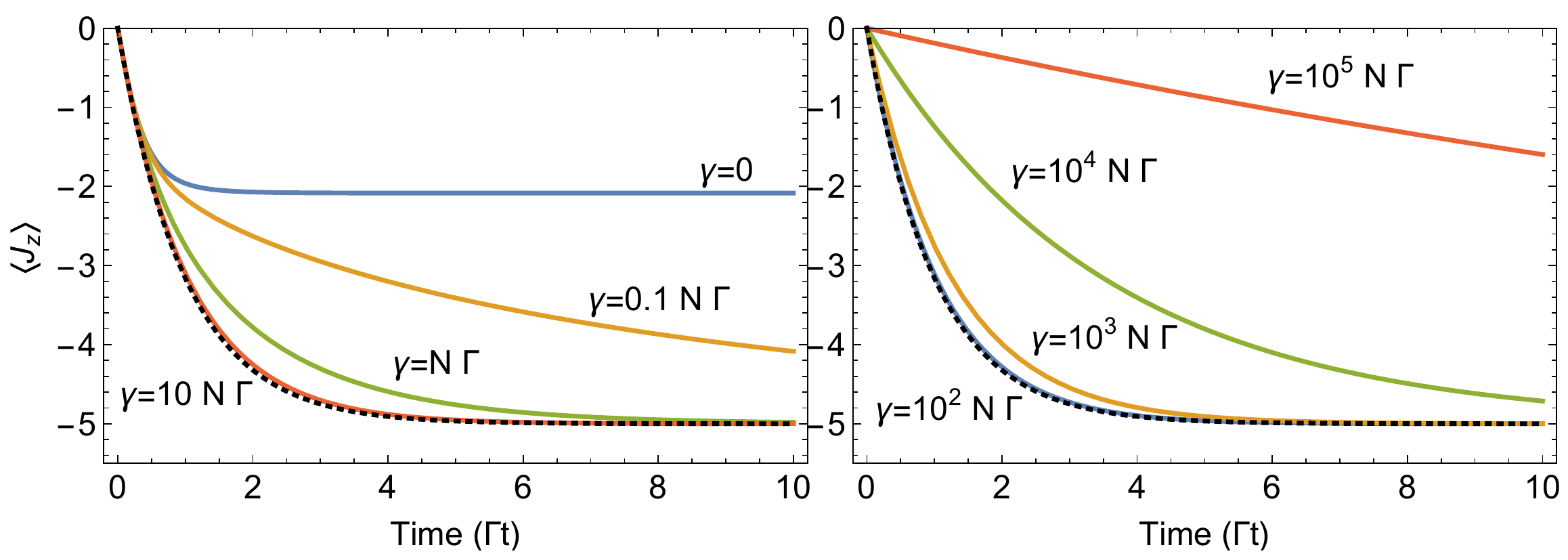}
\caption{Simulations of the magnetization expectation value $\langle J_z(t)\rangle$ for the full spin-cavity master equation in Eq.~\eqref{eq:tc-me} with the addition of a local dephasing dissipator for $N=10$ spins in a maximally mixed initial state, and a cavity truncated to 4 levels initialized in the ground state. Evolution under the 1st order average dissipator is shown as the dotted black line in both figures. 
}
\label{fig2}
\end{center}
\end{figure*}

We numerically compared evolution under $\overline{\2D}_1$ to both the cavity cooling with local dephasing spin master equation in Eq.~\eqref{eq:cct2} (Fig.~\ref{fig1}), and to the full spin-cavity evolution under the Tavis-Cummings master equation with cavity dissipation and local spin dephasing (Fig.~\ref{fig2}):
\begin{equation}
\frac{d}{dt}\rho(t) = -i \left[H_{\text{TC}},\rho(t)\right]
	+\left(\2D_c+\2D_{T_2}\right)\rho(t).
\label{eq:full-me-t2-diss}
\end{equation}
All simulations were done using the \emph{QuantumUtils for Mathematica} package~\cite{QU} in the superoperator representation for the system dynamics~\cite{WoodTN}, using an $N$-spin totally symmetric subspace matrix representation of the SU(4) algebra.
In Fig.~\ref{fig1} we simulated cavity cooling with local dephasing for $N=10$ spins, and $N=100$ spins with dephasing rates of $\gamma = \lambda\, N\, \Gamma$ with $\lambda = 0, 0.1, 1, 10$. In both cases we see that for $\gamma=0$ we have cavity cooling alone and population is trapped. For $\gamma>0$ we break the SU(2)  symmetry and achieve cooling to to the ground state, at a rate that increases with $\gamma$.  For $\gamma= 10 N \Gamma$ we have good agreement with the first order average dissipator expansion $\overline{\2D}_1$. This is expected as Eq.~\eqref{eq:jz-exp} is only a valid approximation of the true dynamics when the dephasing strength $\gamma$ is sufficiently strong to disregard higher order terms in the Magnus expansion where terms that do not commute with the dissipator rapidly average to zero. In practice this corresponds to requiring $\gamma > \Gamma N$. The parameter $C=\Gamma N/\gamma$ is also called the \emph{cooperativity} of the spin ensemble, and hence the condition for the validity of lowest order Magnus approximation is that $C<1$. For example, in a typical X-band pulsed ESR setup reasonable values for cavity dissipation and single spin coupling are $\kappa/2\pi = 1$ MHz and $g/2\pi = 1$ Hz~\cite{Benningshof:2012a}, and the Markovian condition for cavity cooling is satisfied for sample sizes $N < 10^{12}$~\cite{WoodCC}. For these parameters a single spin dephasing rate of $\gamma > 4$ MHz would satisfy $C<1$. We note that the relaxation curve is bi-exponential for $C>1$. Once the initial collective cooling at rate $\Gamma$ is saturated the long time rate is limited by the local dephasing rate.

In Fig.~\ref{fig2} we simulated for $N=10$ spins and a cavity truncated to 4 levels with values of the spin-cavity coupling of $g=100$, cavity dissipation rate $\kappa =4g^2= 4\times10^4$, and spin dephasing rate $\gamma = \lambda\, N$ with $\lambda = 0, 0.1, 1, 10, 10^2, 10^3, 10^4, 10^5$. The values of $g$ and $\kappa$ were chosen to satisfy the Markovian condition $\kappa \gg g\sqrt{N}$ for $N=100$, while giving an effective spin cavity cooling rate of $\Gamma = 4g^2/\kappa = 1$. In addition, the strong cavity dissipation rate allows us to truncate the cavity to low dimension. We find that the spin-cavity master equation is in agreement with the spin cavity cooling master equation for dephasing rates up to $10 N \Gamma$, however as the dephasing rate increases beyond the collective cavity dissipation rate, the cooling rate begins to slow down. Following the derivation in~\cite{WoodCC}, this can be incorporated by a Lorentzian cavity cooling rate $\Gamma= 4g^2\kappa /(\kappa^2+4\Delta^2)$, where $\Delta$ is a parameter that depends the the physical mechanism that gives rise to the local dephasing parameter $\gamma$. For example, in the simplest case of $N=2$, this is the expression where the dephasing mechanism arrises due to the spins being tuned to $\pm\Delta/2$ away from resonance with the cavity.

The reduction in cooling rate in the strong dephasing regime is because the master equation in Eq.~\eqref{eq:cct2} is no longer an honest description of the dynamics: If we consider the full spin-cavity master equation in Eq.~\eqref{eq:full-me-t2-diss}, then as the dephasing rate increases past the collective coupling strength this will suppress the TC exchange interaction for each individual spin with the cavity. To see this we note that in the dissipative interaction frame of $\2D_{T_2}$, the spin cavity master equation in Eq.~\eqref{eq:full-me-t2-diss} has no time dependent piece (See Appendix~\ref{app:magnus}). Hence if we make an imaginary-time transformation and perform a first order Magnus expansion as done in Appendix~\ref{app:localt1} the exchange interaction will be completely averaged out to zero. Thus we find in terms of the full spin-cavity interaction the local-$T_1$ cooling interaction enters at higher-order in the Magnus expansion.

\section{Discussion}
The method we have described shows that cavity cooling techniques can be used to drive an ensemble system into the ground state by coupling to a cavity in the presence of a local dephasing on the spin system. This techniques could prove useful for initialization an ensemble spin system in a highly pure state by short circuiting its thermal relaxation.
There remains an opportunity for yet more efficient methods and the approach we have introduced for computing the cooling rates could be used more broadly. 
In addition the method for dissipative perturbation theory that we developed to solve the cooling master equation should be useful for other systems where a dissipative term is dominant.

Dephasing is always present in a real physical system and this proposal uses it as a resource for dissipative state engineering. Depending on the physical system there are many possible mechanisms that give rise to dephasing. For spins systems it may  arises due to inhomogeneous static fields across the ensemble, which could be engineered by introducing gradient fields. For systems such as defects in silicon and nitrogen vacancy centres in diamond it could arise due to hyperfine coupling between each electron and a local nuclear spin or spin bath, and the strength of these interactions can be engineered to a degree using decoupling sequences. 

Our calculation considered the case where the spin-ensemble was on-resonance with the cavity and the resulting cavity cooling dissipator can be thought of in terms of the Purcell effect. One could also consider the side-band cooling approach as detailed in \cite{WoodCC} by introducing a drive term on the spins to target a side-band of the resonator. In that case the magnetization of the spins under cooling accumulates in the $J_x$ basis (for a $J_x$ drive term) rather than the $J_z$ basis of the static field. In this situation the dephasing must also happen in the $J_x$ basis to achieve cooling to the $J_x$ ground state. In practice this could be engineered by using a gated protocol where a single cooling step consists of: side-band cavity cooling for a time $t_{cc}$, applying a collective rotation swapping the $J_x$ and $J_z$ eigenstates, dephasing for time $t_{T_2}$ in the $J_z$ basis, then applying the inverse collective rotation to rotate back to the $J_x$ basis. This cooling step can then be repeated to form a discretized cavity cooling cycle with dephasing to reach the true ground state.

\begin{acknowledgements}
This work was supported by the Canadian Excellence Research Chairs (CERC) program, and the Natural Sciences and Engineering Research Council of Canada (NSERC) Discovery and CREATE programs.
\end{acknowledgements}


\newpage
\appendix

\section{SU(4) Representation of Collective Spin Dissipators}
\label{app:su4}

We now outline how to represent the dissipators for an ensemble of $N$ two-dimensional subsystems in terms of the generators of SU(4). Rather than explicitly derive this in terms of irreducible representations, we simply give a constructive method for representing each of the generators of SU(4) in terms of superoperators acting on the $N$ subsystems. This approach has been used by \cite{Hartmann2012,Minghui2013} and we use their notation for the SU(4) algebra here. 

There are 15 generators for SU(4), which each belong to one of 6 SU(2) subalgebras. Let $\1 O = \{\2 Q, \Sigma, \2 M, \2 N, \2 U, \2 V\}$ be the set of subalgebra operators. For $\2 O\in \1 O$ we have 
\[
[\2 O_+, \2 O_-] = 2 O_3, \qquad [\2 O_3,\2 O_\pm] = \pm \2 O_\pm.
\]
Also the pairs of operators $(\2Q, \Sigma)$,  $(\2 M, \2 N)$, and $(\2 U, \2V)$ each commute:
\[
[\2 Q_\alpha, \Sigma_\beta] 	
=	[\2 M_\alpha, \2 N_\beta] 
= 	[\2 U_\alpha, \2 V_\beta] 		=	 0	\quad\forall \alpha,\beta\in\{\pm,3\}.
\]
The remainder of the SU(4) commutation relations are shown in Table 1. Note that only 15 of these operators are linearly independent. In particular 
$\2 N_3 = \2 Q_3 + \Sigma_3 - \2 M_3$, 
 $\2 U_3 =  \2 M_3 -\Sigma_3$ and
$\2 V_3 = \2 Q_3 - \2 M_3$.

\begin{table}[htp]
\begin{center}
\[
\begin{array}{|c||ccc|ccc||}
\hline
	 	& 	\2M_{+}		&	\2M_{-}		&	\2M_{3}		& 	\2N_{+}	&	\2N_{-}	&	\2N_{3}	\\
\hline\hline
\2 Q_+ 	& 	0			&   	-\2 V_{+} 		& -\frac12\2 Q_{+} 	& 	0			& 	\2U_{+}		& 	-\frac12\2 Q_{+}	\\
\2 Q_- 	& 	\2V_{-}		& 	0			& \frac12\2 Q_{-}	& 	-\2U_{-}		& 	0			& 	\frac12\2 Q_{-}		\\
\2 Q_3 	& \frac12 \2 M_{+} 	&-\frac12 \2 M_{-} 	& 	0			& \frac12 \2 N_{+} 	&-\frac12 \2 N_{-} 	& 	0				\\
\hline
\Sigma_+ 	& 	0			&   	\2 U_{-} 		&-\frac12 \Sigma_{+} & 	0			& 	-\2V_{-}		&-\frac12 \Sigma_{+}	\\
\Sigma_- 	& 	-\2U_{+}		& 	0			& \frac12 \Sigma_{-}	& 	\2V_{+}		& 	0			& \frac12 \Sigma_{-}		\\
\Sigma_3 	& \frac12 \2 M_{+} 	&-\frac12 \2 M_{-} 	& 	0			& \frac12 \2 N_{+} 	&-\frac12 \2 N_{-} 	& 	0				\\
\hline	
\end{array}
\]
\[
\begin{array}{|c||ccc|ccc||}
\hline
	 	& 	\2U_{+}		&	\2U_{-}		&	\2U_{3}		& 	\2V_{+}	&	\2V_{-}	&	\2V_{3}	\\
\hline\hline
\2 Q_+ 	& 	0			&   	-\2 N_{+} 		& -\frac12\2 Q_{+} 	& 	0			& 	\2M_{+}		& 	-\frac12\2 Q_{+}	\\
\2 Q_- 	& 	\2N_{-}		& 	0			& \frac12\2 Q_{-}	& 	-\2M_{-}		& 	0			& 	\frac12\2 Q_{-}		\\
\2 Q_3 	& \frac12 \2 U_{+} 	&-\frac12 \2 U_{-} 	& 	0			& \frac12 \2 V_{+} 	&-\frac12 \2 V_{-} 	& 	0				\\
\hline
\Sigma_+ 	& 	-\2 M_{+}		&   	0 			& \frac12 \Sigma_{+} & 	\2N_{+}		& 	0			&	\frac12 \Sigma_{+}	\\
\Sigma_- 	& 	0			& 	\2M_{-}		& -\frac12 \Sigma_{-}	& 	0			& 	-\2N_{-}		& -\frac12 \Sigma_{-}		\\
\Sigma_3 	& -\frac12 \2 U_{+} 	&\frac12 \2 U_{-} 	& 	0			& -\frac12 \2 V_{+} 	&\frac12 \2 V_{-} 	& 	0				\\
\hline	
\end{array}
\]

\[
\begin{array}{|c||ccc|ccc||}
\hline
	 	& 	\2U_{+}		&	\2U_{-}		&	\2U_{3}		& 	\2V_{+}	&	\2V_{-}	&	\2V_{3}	\\
\hline\hline
\2 M_+ 	& 	0			&   	- \Sigma_{+} 	& -\frac12\2 M_{+} 	& 	\2 Q_{+}		& 	0			& 	\frac12\2 M_{+}	\\
\2 M_- 	& 	 \Sigma_{-}		& 	0			& \frac12\2 M_{-}	& 	0			& 	-\2 Q_{-}		& 	-\frac12\2 M_{-}		\\
\2 M_3 	& \frac12 \2 U_{+} 	&-\frac12 \2 U_{-} 	& 	0			& -\frac12 \2 V_{+} 	& \frac12 \2 V_{-} 	& 	0				\\
\hline
\2 N_+ 	& 	-\2 Q_{+}		&   	0 			& \frac12 \2 N_{+} 	& 	0			& 	 \Sigma_{+}	&	-\frac12 \2 N_{+}	\\
\2 N_- 	& 	0			& 	\2Q_{-}		& -\frac12 \2 N_{-}	& 	-\Sigma_{-	}	& 	0			& 	\frac12 \2 N_{-}		\\
\2 N_3 	& -\frac12 \2 U_{+} 	&\frac12 \2 U_{-} 	& 	0			& \frac12 \2 V_{+} 	&-\frac12 \2 V_{-} 	& 	0				\\
\hline	
\end{array}
\]
\end{center}
\caption{Commutation relations for SU(4) algebra.The table entry is the value of the commutator $\left[\2O_{\text{row}},\2O_{\text{col}}\right]$ where $\2O_{\text{row}}$ and $\2O_{\text{col}}$ are the corresponding operators in the same row and column of the table labels respectively.}
\label{default}
\end{table}

The SU(4) generators may be given an explicit matrix representation in terms of superoperators acting on the vectorized density matrices of $N$ two-dimensional subsystems. This is analogous to how the spin operators, or SU(2) algebra, can be given a $(2J+1)$-dimensional matrix representation in terms of the spin operators acting on a spin-$J$ particle. Recall that for a quantum system with Hilbert space $\2X\cong \bb C^d$, density matrices are square matrices corresponding to linear maps $\rho:\2X\rightarrow\2X$. We can vectorize density matrices by stacking the columns together to form a column vector $\dket{\rho}\in \2X\otimes\2X$. Superoperators are then the linear operators which act on vectorized density matrices: $\2S\dket{\rho} = \dket{\rho^\prime}$. A linear map $\2E(\rho)$ given by $\2E(\rho) = A\rho B^\dagger$ can be written as a superoperator as $\2S_{\2E}= B^*\otimes A$ where $^*$ denotes complex conjugation~\cite{WoodTN}.
A superoperator acting on $N$ subsystems can be written as  $\2S_{\2E}=\sum_j B^{(j)*}\otimes A^{(j)}$, and in particular the SU(4) operators may be represented in terms of the single spin Pauli operators $\sigma_{\pm}^{(j)}, \sigma_z^{(j)}, E_\pm^{(j)}=\frac12(\id+\sigma_z^{(j)})$, and collective spin operators $J_z= \sum_{j=1}^N \sigma_z^{(j)}/2$, $J_\pm= \sum_{j=1}^N \sigma_\pm^{(j)}$ as follows:
\begin{align*}
\2 Q_{\pm}	&=	\sum_{j=1}^N\left( \sigma_\pm^{(j)}\otimes \sigma_\pm^{(j)} \right),	&
\2 Q_{3} 		&=  \frac12\left( \id\otimes J_z+J_z\otimes\id\right), 	\\
 \Sigma_{\pm} 	&=  	\sum_{j=1}^N\left(\sigma_\mp^{(j)} \otimes \sigma_\pm^{(j)}\right),	&
\Sigma_{3}	&=  	\frac12\left( \id\otimes J_z-J_z\otimes\id\right), 	\\
\2 M_{\pm}	&=  	\sum_{j=1}^N\left( E_+^{(j)}\otimes \sigma_\pm^{(j)} \right),	&
\2 M_{3}		&=  	\frac 12\sum_{j=1}^N\left( E_+^{(j)}\otimes \sigma_z^{(j)} \right),	\\
\2 N_{\pm}	&=	\sum_{j=1}^N\left( E_-^{(j)}\otimes \sigma_\pm^{(j)}	 \right),	&
\2 N_{3}		&=	\frac 12\sum_{j=1}^N\left( E_-^{(j)}\otimes \sigma_z^{(j)} \right),	\\
\2 U_{\pm}	&=	\sum_{j=1}^N\left( \sigma_\pm^{(j)}\otimes E_+^{(j)} \right),	&
\2 U_{3}		&=	\frac 12\sum_{j=1}^N\left(\sigma_z^{(j)}\otimes E_+^{(j)} \right),	\\
\2 V_{\pm}		&=	\sum_{j=1}^N\left( \sigma_\pm^{(j)}\otimes E_-^{(j)} \right),	&
\2 V_{3}		&=	\frac 12\sum_{j=1}^N\left( \sigma_z^{(j)}\otimes E_-^{(j)} \right).
\end{align*}

The utility of this approach is that we can express many useful open system dynamics of collective spins in this representation, and it allows us to analytically compute certain properties that may be difficult otherwise.
In particular local $T_2$ and $T_1$ process dissipators may be expressed as
\begin{align}
 \sum_{j=1}^N D[\sigma_z^{(j)}/2] 
	&= \2M_3 - \frac12 \2Q_3 - \frac12 \Sigma_3 - \frac{N}{4}\2I, 	
	\label{eq:loc-t2}\\
 \sum_{j=1}^N D[\sigma_\pm^{(j)}] 
	&= 2 \2Q_\pm  +2\2Q_3 	-N\, \2I,
	\label{eq:loc-t1}
\end{align}
where $\2I=\id$ is the SU(4) identity operator. Collective $T_2$ and $T_1$ process may be expressed as
\begin{align}
D\left[J_z\right] 
	&= - 2\Sigma_3^2,	
	\label{eq:col-t2}\\
D[J_\pm] 
	&=  (\2U_\pm + \2V_\pm)(\2M_\pm + \2N_\pm)
	\nonumber\\&\qquad
		-\frac{1}{2}(\2U_\mp + \2V_\mp)(\2U_\pm + \2V_\pm)
	\nonumber\\&\qquad
		-\frac{1}{2}(\2M_\mp + \2N_\mp)(\2M_\pm + \2N_\pm).
	\label{eq:col-t1}
\end{align}

\section{Derivation of Local $T_1$ Dissipation}
\label{app:localt1}

Consider a system with a cavity cooling dissipator, and a local $T_2$ dissipator.
\begin{align}
\2D_{cc} &= \Gamma (1+\overline{n})D[J_-] + \Gamma\, \overline{n} \,D[J_+],	\\
\2D_{T_2}	&= \gamma \sum_{j=1}^N D\left[\sigma_z^{(j)}/2\right]. 
\end{align}
The superoperators for these dissipators are given in terms of SU(4) generators by Eq.~\eqref{eq:col-t1} and \eqref{eq:loc-t2} respectively.
We now consider the effective cavity cooling dissipator in the interaction frame of the $T_2$ dissipator: 
\begin{equation}
\tilde{\2D}_{cc}(t) =
\Gamma (1+\overline{n})\tilde{D}[J_-](t) + \Gamma\,\overline{n} \tilde{D}[J_+](t),	
\end{equation}
where
\begin{equation}
 \tilde{D}[J_\pm](t) = e^{t \2D_{T_2}} D[J_\pm]e^{-t \2D_{T_2}}.
\end{equation}
We may expand this using the BCH exansion:
\begin{equation}
 \tilde{D}[J_\pm](t) = \sum_{k=0}^\infty \frac{t^k}{k!}  \2C_k\big[ D[J_\pm]\big],
\end{equation}
where $\2C_k\big[D[J_\pm]\big]$ are nested commutator terms with
\begin{subequations}
\begin{align}
\2C_0\big[D[J_\pm]\big]	& = D[J_\pm],	\\
\2C_1\big[D[J_\pm]\big]	& = \big[\2D_{T_2}, D[J_\pm]\big],	\\
\2C_k\big[D[J_\pm]\big]	& = \Big[\2D_{T_2}, \2C_{k-1}\big[D[J_\pm]\big]\Big].
\end{align}
\end{subequations}
We may compute the commutator terms of the BCH expansion using the SU(4) algebra.
To begin, we have
\begin{subequations}
\begin{align}
\big[ \2D_{T_2},\,  \2M_\pm \big]
	&= \pm\frac{\gamma}{2}\2M_\pm, 
&\big[ \2D_{T_2},\,  \2N_\pm \big]
	&= \mp \frac{\gamma}{2} \2N_\pm,	\\
\big[ \2D_{T_2},\,  \2U_\pm \big]
	&= \pm \frac{\gamma}{2} \2U_\pm,
&\big[ \2D_{T_2},\,  \2V_\pm \big]
	&= \mp \frac{\gamma}{2} \2V_\pm.
\end{align}
\end{subequations}
Hence 
\begin{align}
\Big[ \2D_{T_2},\,  D[J_\pm]  \Big]
	&= \frac{1}{2}\bigg(
		\big[ \2D_{T_2},\, 
			(\2U_\pm + \2V_\pm)(\2M_\pm + \2N_\pm)\big]
	\nonumber\\&\qquad
		+\big[\2D_{T_2},\, 
			(\2M_\pm + \2N_\pm)(\2U_\pm + \2V_\pm)\big] 
	\nonumber\\&\qquad
	 	-\big[ \2D_{T_2},\,
			(\2U_\mp + \2V_\mp)(\2U_\pm + \2V_\pm)\big]
	\nonumber\\&\qquad
		-\big[ \2D_{T_2},\,
			(\2M_\mp + \2N_\mp)(\2M_\pm + \2N_\pm)\big]
		\bigg)
	\nonumber\\
	= \pm\frac{\gamma}{2}&
		\bigg[
		\big(\2M_\pm\2U_\pm+\2U_\pm\2M_\pm 
		- \2V_\mp\2U_\pm-\2N_\mp\2M_\pm \big)
	\nonumber\\
	- &\big(	
		\2N_\pm\2V_\pm
		+\2V_\pm\2N_\pm
		- \2U_\mp\2V_\pm
		- \2M_\mp\2N_\pm
	\big)
		\bigg].
\end{align}
Define superoperators
\begin{align}
A_\pm &=\frac{1}{2}\big(
		\2M_\pm\2U_\pm
		+\2U_\pm\2M_\pm 
		- \2V_\mp\2U_\pm
		-\2N_\mp\2M_\pm \big),
		\\
B_\pm &= \frac{1}{2}\big(
		\2N_\pm\2V_\pm
		+\2V_\pm\2N_\pm
		- \2U_\mp\2V_\pm
		- \2M_\mp\2N_\pm\big),
\end{align}
then we may write
\begin{align}
\Big[ \2D_{T_2},\,  D[J_\pm]  \Big]
	&=  \pm\gamma A_\pm \mp\gamma B_\pm.
\end{align}
If we then take the commutator of $A_\pm$ and $B_\pm$ with the $T_2$ dissipator we find
\begin{align}
\big[ \2D_{T_2},\, A_\pm \big]
	&= \pm\gamma A_\pm,
&\big[ \2D_{T_2},\, B_\pm \big]
	&= \mp\gamma B_\pm.
\end{align}
Hence for $k\ge 1$ we have that the nested commutator terms are given by
\begin{equation}
\2C_k\big[D[J_\pm]\big] = (\pm \gamma)^k  A_\pm + (\mp \gamma)^k  B_\pm.
\end{equation}
Thus the interaction frame dissipator terms are given by
\begin{align}
\tilde{D}[J_\pm](t)
	&= \sum_{k=0}^\infty \frac{t^k}{k!}\2C_k\big[D[J_\pm]\big] \nonumber\\
	 &= D[J_\pm] - A_\pm - B_\pm
	 \nonumber\\&\qquad
	 +\sum_{k=0}^\infty \frac{(\pm\gamma t)^{k}}{k!} A_\pm
	 +\sum_{k=0}^\infty \frac{(\mp\gamma t)^{k}}{k!} B_\pm \nonumber\\
	 &= D[J_\pm] 
	 +(e^{\pm\gamma t}-1) A_\pm
	 +(e^{\mp\gamma t}-1) B_\pm.
\end{align}

If we perform a change of variables to imaginary time the interaction frame operator $G(\tau)=\tilde{D}[J_\pm](i \tau)$ is periodic with period $T=2\pi/\gamma^{-1}$. The first order Magnus term over this period is the time-independent piece of the dissipator:
\begin{equation}
\overline{\2D}_1 =  \overline{G}_1 = \Gamma(1+\overline{n})\,\overline{G}_{-} + \Gamma\,\overline{n}\,\overline{G}_{+},
\label{eq:app-1st-diss}
\end{equation}
where
\begin{align}
\overline{G}_{\pm}
	&= D[J_\pm] - A_\pm - B_\pm  \nonumber\\
	&=\frac{1}{2}\Big(
			\2U_\pm\2N_\pm + \2N_\pm\2U_\pm
			+\2M_\pm\2V_\pm + \2V_\pm\2M_\pm
			\Big)
	\nonumber\\&\qquad
		-\frac{1}{2}\Big(
			\2U_\mp\2U_\pm + \2V_\mp\2V_\pm
			+\2M_\mp\2M_\pm + \2N_\mp \2N_\pm
		\Big).
		\label{eq:sec-diss}
\end{align}

\subsection{$J_z$ Expectation Value}
\label{app:jz}
We are interested in computing the evolution of the expectation value for the $J_z$ operator for an arbitrary initial state for dynamics described by the first order average dissipator in Eq.~\eqref{eq:app-1st-diss}. We do this by solving
\begin{align*}
\langle J_z(t)\rangle
	&= \dbradket{J_z}{\rho(t)}	
	=  \dbradket{J_z}{\tilde{\rho}(t)}	\\
	&=   \dbra{J_z}e^{t \overline{\2D}_1}\dket{\tilde{\rho}(0)}	
	=   \dbra{\tilde{\rho}(0)}e^{t \overline{\2D}_1^\dagger}\dket{J_z},
\end{align*}
where $\tilde{J}_z(t) = J_z(t)$ as $[J_z,\2D_{T_2}]=0$.
The adjoint dissipator is given by
\begin{equation}
\overline{\2D}_1^\dagger =  \overline{G}_1^\dagger = \Gamma(1+\overline{n})\overline{G}_{-}^\dagger + \Gamma\,\overline{n}\overline{G}_{+}^\dagger,
\label{eq:app-1st-ad-diss}
\end{equation}
where
\begin{align}
\overline{G}_{\pm}^\dagger
	&=\frac{1}{2}\Big(
			\2U_\mp\2N_\mp + \2N_\mp\2U_\mp
			+\2M_\mp\2V_\mp + \2V_\mp\2M_\mp
			\Big)
	\nonumber\\&\qquad
		-\frac{1}{2}\Big(
			\2U_\mp\2U_\pm + \2V_\mp\2V_\pm
			+\2M_\mp\2M_\pm + \2N_\mp \2N_\pm
		\Big), 
\end{align}
and we have used $\2M_\pm^\dagger = \2M_\mp$ and similarly for $\2N_\pm, \2U_\pm, \2V_\pm$. To compute the terms of $\overline{G}^\dagger \dket{J_z}$ we have that
\begin{align*}
(\2U_\mp \2N_\mp +\2N_\mp\2U_\mp)\dket{J_z}
	&=
		\sum_{i,j=1}^N \bigg( \dket{E_{+}^{(j)}  \sigma_\mp^{(i)} J_z E_{-}^{(i)}  \sigma_\pm^{(j)}}
	\nonumber\\&\qquad
			+ \dket{\sigma_\mp^{(i)} E_{+}^{(j)}   J_z  \sigma_\pm^{(j)}  E_{-}^{(i)}}
	\bigg) \nonumber\\
	&=
		\sum_{i,j=1}^N \bigg( \dket{E_{+}^{(j)}  \sigma_\mp^{(i)}E_{-}^{(i)}   J_z \sigma_\pm^{(j)}}
	\nonumber\\&\quad
			+ \dket{\sigma_\mp^{(i)} J_z E_{+}^{(j)}    \sigma_\pm^{(j)}  E_{-}^{(i)}}.
\end{align*}
Hence
\begin{align*}
(\2U_{-} \2N_{-} +\2N_{-} \2U_{-})\dket{J_z}
	&= \sum_{i,j=1}^N\dket{ \sigma_{-}^{(i)}\, J_z \, \sigma_{+}^{(j)} \, E_{-}^{(i)}}	\nonumber\\
	&= \sum_{i,j=1}^N\bigg(  
			\dket{\left[\sigma_{-}^{(i)}, J_z\right] \sigma_{+}^{(j)} \, E_{-}^{(i)}}
	\nonumber\\&\qquad
			+ \dket{J_z \sigma_{-}^{(i)}\sigma_{+}^{(j)} \, E_{-}^{(i)}}	
			\bigg)\nonumber\\
	&=\sum_{j=1}^N\dket{\left(J_z +\id \right) 
			 E_{-}^{(j)} }	\nonumber\\
	&=\dket{\left(J_z +\id \right) \left( \frac{N}{2}\id - J_z \right)	},
\end{align*}
and
\begin{align*}
(\2U_{+} \2N_{+} +\2N_{+} \2U_{+})\dket{J_z}
	&= \sum_{i,j=1}^N \dket{ E_{+}^{(j)} \, \sigma_{+}^{(i)} \, J_z\, \sigma_{-}^{(j)}} \nonumber\\
	&= \sum_{i,j=1}^N  \dket{E_{+}^{(j)} \, \sigma_{+}^{(i)} \sigma_{-}^{(j)} 
		\left(J_z - \id\right)} \nonumber\\
	&=  \dket{\left(\frac{N}{2}\id + J_z\right)
		\left(J_z - \id\right)}
\end{align*}
where we have made use of the relations
\begin{align*}
E_\pm E_\pm &= E_\pm,\\
E_\pm E_\mp &=0,\\
E_\pm \sigma_\pm	&= \sigma_\pm E_\mp = \sigma_\pm \\
E_\mp \sigma_\pm	&= \sigma_\pm E_\pm =0.
\end{align*}
Thus we have
\[
(\2U_\mp \2N_\mp +\2N_\mp\2U_\mp)\dket{J_z} 
 = \dket{\left(\frac{N}{2}\id \mp J_z\right)\left(J_z \pm \id\right)}.
\]
Similarly one can show 
\[
(\2V_\mp \2M_\mp +\2M_\mp\2V_\mp)\dket{J_z} 
 = \dket{\left(\frac{N}{2}\id \mp J_z\right)\left(J_z \pm \id\right)}.
\]
For the other terms we have
\begin{align*}
\2U_{\mp} \2U_{\pm} \dket{J_z}
	&=\sum_{i,j=1}^N \dket{E_+^{(j)} E_+^{(i)} J_z  \sigma_{\mp}^{(i)} \sigma_{\pm}^{(j)}} \nonumber\\
	&= \sum_{i,j=1}^N \dket{J_z  \, E_+^{(j)} \, E_+^{(i)} \sigma_{\mp}^{(i)} \sigma_{\pm}^{(j)}}
\end{align*}
and hence
\begin{align*}
\2U_{-} \2U_{+} \dket{J_z}
	&= \sum_{i,j=1}^N \dket{J_z  \, E_+^{(j)} \, E_+^{(i)} \sigma_{-}^{(i)} \sigma_{+}^{(j)}}\\
	&= 0,\\
\2U_{+} \2U_{-} \dket{J_z}
	&= \sum_{i,j=1}^N \dket{J_z  \, E_+^{(j)} \,  \sigma_{+}^{(i)} \sigma_{-}^{(j)}}\\
	&= \sum_{j=1}^N \dket{J_z  \, E_+^{(j)}}\\
	&=  \dket{J_z \left(\frac{N}{2}\id + J_z\right)}.
\end{align*}
Similarly
\begin{align*}
\2M_{-} \2M_{+} \dket{J_z} &= 0, \\
\2M_{+} \2M_{-} \dket{J_z} &= \dket{J_z \left(\frac{N}{2}\id + J_z\right)}, \\
\2V_{-} \2V_{+} \dket{J_z} &= \dket{J_z \left(\frac{N}{2}\id - J_z\right)},  \\
\2V_{+} \2V_{-} \dket{J_z} &=  0,\\
\2N_{-} \2N_{+} \dket{J_z} &= \dket{J_z \left(\frac{N}{2}\id - J_z\right)},  \\
\2N_{+} \2N_{-} \dket{J_z} &=  0.
\end{align*}
Thus we have
\begin{align}
\overline{G}_{\pm}^\dagger \dket{J_z}
	&=  \dket{\left(\frac{N}{2}\id \mp J_z\right)J_z}
		\pm  \dket{\left(\frac{N}{2}\id \mp J_z\right)}
		\nonumber\\&\qquad
		-\dket{\left( \frac{N}{2}\id \mp J_z\right)J_z} \nonumber\\
	& = - \dket{J_z}  \pm  \frac{N}{2}\dket{\id}.
\end{align}
Next we need to evaluate $\overline{G}_{\pm}^\dagger\dket{\id}$. 
We have that
\begin{align*}
\2M_+\dket{\id} &= 0,			&	\2M_-\dket{\id} &= \dket{J_-}, \\
\2U_+\dket{\id} &= 0, 		&	\2U_-\dket{\id} &= \dket{J_+}, \\
\2N_+\dket{\id} &= \dket{J_+}, 	&	\2N_-\dket{\id} &= 0, \\
\2V_+\dket{\id} &= \dket{J_-},	&	\2V_-\dket{\id} &= 0,
\end{align*}
and so
\begin{align*}
\overline{G}_{+}^\dagger\dket{\id}
	&= \frac{1}{2}\Big(\2N_- \dket{J_+} +\2V_- \dket{J_-}\big)
		\nonumber\\&\qquad
		-\frac{1}{2}\Big(\2N_- \dket{J_+} +\2V_- \dket{J_-}\big)
		\\
	& = 0, \\
\overline{G}_{-}^\dagger\dket{\id}
	&= \frac{1}{2}\Big(\2M_+ \dket{J_-} +\2U_+ \dket{J_+}\big)
	\nonumber\\&\qquad
		-\frac{1}{2}\Big(\2M_+ \dket{J_-} +\2U_+ \dket{J_+}\big)
	 \\&= 0.
\end{align*}
Finally we may put this all together to obtain
\begin{align}
\overline{G}^\dagger\dket{J_z}
	&= \Gamma\left((1+\overline{n})\overline{G}_-^\dagger
		+\overline{n}\,\overline{G}_+^\dagger
		\right) \dket{J_z}
	 \nonumber\\
	&= \Gamma(1+\overline{n})(- \dket{J_z}  -  \frac{N}{2}\dket{\id})
	\nonumber\\ &\qquad
		+\Gamma\,\overline{n}(- \dket{J_z}  +  \frac{N}{2}\dket{\id})
	\nonumber\\
	&= -\Gamma(1+2\overline{n})\left[ 
		\dket{J_z}  
		+\left(
			\frac{N}{2+4\overline{n}}
			\right)\dket{\id}
		\right],
\end{align}
and in general for $k\ge 1$
\begin{align}
\overline{G}^k \dket{J_z}
 & =  (-1)^k\Gamma^k(1+2\overline{n})^k \bigg[
 		\dket{J_z}
 	+\left(
		\frac{N}{2+4\overline{n}}
		\right)\dket{\id}
		\bigg].
  \nonumber\\
\end{align}
Hence we have
\begin{align}
e^{t \overline{G}}\dket{J_z}
	&= \sum_{k=0}^\infty \frac{t^k}{k!}
		\overline{G}^k\dket{J_z}
		\nonumber\\
  	&= e^{-t\,\Gamma(1+2\overline{n})}\dket{J_z}
		-\left(\frac{N\left(1-e^{-t\,\Gamma(1+2\overline{n})}\right)}{2+4\overline{n}}
		\right)\dket{\id}.
\end{align}
Thus for an arbitrary initial state $\rho$ the expectation value of $J_z$ under this evolution is given by
\begin{align}
\langle J_z(t)\rangle
	&=e^{-t\,\Gamma(1+2\overline{n})}\langle J_z(0)\rangle
			\nonumber\\&\qquad
		-\left(1-e^{-t\,\Gamma(1+2\overline{n})}\right)
		\left(
		\frac{N}{2+4\overline{n}}
		\right).
\end{align}
Hence the effective dynamics are described by an exponential decay process with decay rate 
\begin{equation}
T_{1}=\frac{1}{\Gamma(1+2\overline{n})}
\end{equation}
to an equilibrium state with magnetization
\begin{equation}
\langle J_z\rangle_{eq} = -\frac{N}{2+4\overline{n}},
\end{equation}
where in the ideal cooling limit of $\overline{n}=0$, this is a $T_1$ process to the ground state of the spin ensemble.

\section{Strong Dephasing Limit}
\label{app:magnus}

In this appendix we outline a model for the cooling results in the presence of strong dephasing in Sec.~\ref{sec:sim}. The cooling dynamics in Fig.~\ref{fig2} shows the spin-ensemble magnetization for evolution under the the full spin-cavity master equation with local dephasing:
\begin{equation}
\frac{d}{dt}\dket{\rho(t)} = \left(\2S_{\text{TC}}+ \2D_c + \2D_{T_2}\right)\dket{\rho(t)},
\label{eq:app-tc-me}
\end{equation}
where $\rho(t)$ is the density matrix for the joint spin-cavity system and $\2S_{\text{TC}}$ is the superoperator for the TC interaction
\begin{equation}
\dket{\rho(t)^\prime}\equiv \2S_{\text{TC}}\dket{\rho(t)} \Longleftrightarrow \rho(t)^\prime \equiv -i[H_{\text{TC}},\rho(t)].
\end{equation}

As described in Sec~\ref{sec:sim}, as the dephasing rate increases beyond the collective cavity dissipation rate the cooling rate begins to slow down. This is because in the strong dephasing regime the spin-cavity interaction term $\2S_{\text{TC}}$ is suppressed. If we consider the Magnus expansion of $\2S_{\text{TC}}$ in the interaction frame defined by $\2D_{T_2}$ we find that there is no secular piece that commutes with the dephasing interaction frame. Let $\2S_\pm^{(j)}$ be the superoperator defined by $ \2S_\pm^{(j)} \dket{\rho_s} \Leftrightarrow -i [\sigma_\pm^{(j)},\rho_s]$. In column stacking convention this is given by
\begin{equation}
\2S_\pm^{(j)} = -i \left(\sigma_{+}^{(j)}\otimes \id^{(j)} - \id^{(j)}\otimes \sigma_{-}^{(j)}\right).
\end{equation}
If we move into the interaction frame of $\gamma \2D[\sigma^{(j)}_z]$ we have
\begin{equation}
\widetilde{\2S}_\pm^{(j)}(t) = -i \left(e^{2t\gamma}\, \sigma_{+}^{(j)}\otimes \id^{(j)} - e^{-2t\gamma}\,\id^{(j)}\otimes \sigma_{-}^{(j)}\right)
\end{equation}
and so all terms of $\widetilde{\2S}_{\text{TC}}(t)$ have time-dependence of $e^{\pm 2t \gamma}$.
We can see here that if we make an imaginary-time transformation and perform a first order Magnus expansion as done in Appendix~\ref{app:localt1} that the exchange interaction will be completely averaged out to zero. Hence in terms of the full spin-cavity interaction the local-$T_1$ cooling interaction enters at higher-order in the Magnus expansion.

\bibliography{references}{}
\bibliographystyle{apsrev4-1}
\end{document}